\documentclass[sigplan,10pt]{acmart}
\renewcommand\footnotetextcopyrightpermission[1]{}
\AtBeginDocument{%
  }

\usepackage{subcaption}
\usepackage{tabularx}
\usepackage{xcolor}
\usepackage{listings}
\usepackage{enumitem}

\definecolor{codekw}{RGB}{83,180,230}

\definecolor{codetype}{RGB}{0,158,115}

\definecolor{codecomment}{RGB}{190,170,0}
\definecolor{codestring}{RGB}{213,94,0}
\definecolor{codelinenum}{RGB}{120,120,120}

\usepackage{multirow}

\lstdefinestyle{polyfetchcpp}{
  language=C++,
  basicstyle=\footnotesize\ttfamily,
  keywordstyle=\bfseries\color{codekw},
  keywordstyle=[2]\bfseries\color{codetype},
  commentstyle=\itshape\color{codecomment},
  stringstyle=\color{codestring},
  numbers=left,
  numberstyle=\tiny\color{codelinenum},
  numbersep=6pt,
  frame=none,
  breaklines=true,
  breakatwhitespace=false,
  columns=fullflexible,
  keepspaces=true,
  showstringspaces=false,
  tabsize=2,
  xleftmargin=1.2em,
  xrightmargin=0em,
  aboveskip=0.2em,
  belowskip=0.1em,
  morekeywords=[2]{File,Page,MarkovModel,Pages,Request,LaunchPolicy,SQLitePolicy,LLMPolicy,BTreePolicy,Requests,Layer,Tensor,TENSOR_LAYOUT,LAYER_START}
}

\setcopyright{none}

\begin{document}
\pagestyle{plain}

\title{\NAME{}: Decoupling Workload Prediction from OS-Controlled Storage Prefetching}


\def\NAME{\textsc{PredictIO}}


\author{Jun Lu}
\email{junlu@tsinghua.edu.cn}
\affiliation{%
  \institution{Tsinghua University}
  \city{Beijing}
  \country{China}
}

\author{Yancan Mao}
\email{maoyancan@hust.edu.cn}
\affiliation{%
  \institution{Huazhong University of Science and Technology}
  \city{Wuhan}
  \country{China}
}

\author{Zhenya Ma}
\email{zhenyama99@gmail.com}
\affiliation{%
  \institution{Tsinghua University}
  \city{Beijing}
  \country{China}
}

\author{Sheng Yue}
\email{yuesh5@mail.sysu.edu.cn}
\affiliation{%
  \institution{Sun Yat-sen University}
  \city{Guangzhou}
  \country{China}
}

\author{Yongjian Fu}
\email{fuyongjian@tsinghua.edu.cn}
\affiliation{%
  \institution{Tsinghua University}
  \city{Beijing}
  \country{China}
}

\author{Ju Ren}
\email{renju@tsinghua.edu.cn}
\affiliation{%
  \institution{Tsinghua University}
  \city{Beijing}
  \country{China}
}

\renewcommand{\shortauthors}{Trovato et al.}






\begin{abstract}
Storage stalls remain a significant source of application latency even on fast storage devices. Prefetching can hide part of this latency by fetching data before it is demanded, but general-purpose operating-system mechanisms typically infer future accesses from recent file offsets. This approach is ineffective when accesses are determined by workload structure, such as recurring sparse accesses during application launch, logical traversal in database scans, or model-layout metadata in LLM runtimes. Existing workload-aware prefetchers usually embed prediction logic inside individual applications and couple execution to application-level I/O paths, making policies difficult to reuse across workloads and systems.

We present \NAME{}, a programmable interface that separates workload prediction from prefetch execution. \NAME{} delivers kernel-observed accesses and application hints to user-space policies, which express predictions as uniform page-range requests. A kernel-space executor validates, bounds, and filters these requests against cached and in-flight pages before issuing asynchronous prefetch I/O. We implement \NAME{} for application launch, SQLite range scans, and LLM model loading. Across these workloads, \NAME{} reduces latency by 5.3--26.3\%, achieves up to 2.99$\times$ speedup for SQLite scans, and reduces LLM loading latency by 8.8--47.9\%. We further show that a common prediction policy can be reused across SQLite and LMDB through lightweight adapters, that an application-integrated RocksDB readahead policy can be migrated while preserving its workload selectivity, and that the execution backend determines how much benefit migration provides for different access patterns and storage devices.
\end{abstract}
\settopmatter{printfolios=true, printacmref=false}
\maketitle

\section{Introduction}
Storage I/O latency remains an important source of application stall time even on fast storage devices. Prefetching hides this latency by loading data into memory before it is demanded. In Linux, the primary general-purpose mechanism is readahead, which fetches nearby file pages based on recently observed file offsets~\cite{wu2007linux,pai2004linux}. This works well for streaming workloads, where future accesses are usually close to recent ones, but it relies on spatial locality in the file-offset stream as a predictor of future access. Many modern workloads violate this offset-locality assumption~\cite{gill2005sarc,griffioen1994reducing,liu2017graphene,li2026optimizations,jia2020vsmt,xing2023maximizing,lu2017wisckey,ahmad2015compaction,zeng2023empirical}. Application launch touches sparse pages across executables, shared libraries, JAR files, and assets~\cite{ryu2023fast,bijlani2021did}; database range scans traverse B-trees according to logical key order rather than file-offset locality~\cite{chen2001improving,chen2002fractal}; and LLM runtimes load model shards according to tensor layouts and execution order~\cite{liu2026accelerating,maurya2025mlp,ren2025characterizing}. In each case, the workload exposes useful structure, but the storage stack ultimately observes only a stream of page accesses. As a result, readahead may fail to identify the right pages early enough, leaving demand I/O on the critical path.

This gap is not merely about more I/O. When future accesses are determined by workload structure rather than offset locality, effective prefetching depends on making prediction signals available before demand occurs. 
At the root of this limitation is a visibility asymmetry: the kernel sees file offsets, while applications and runtimes often possess the information that makes future I/O predictable in advance. Consequently, improving prefetching is not merely a matter of designing a better kernel heuristic. It calls for a common OS abstraction that makes application-visible knowledge available to the prefetching path without embedding workload-specific semantics in the kernel.

Prior work has explored workload-aware prefetching from two broadly complementary directions. Workload-specific prefetchers exploit database semantics or recurring access history to predict future accesses, demonstrating that workload structure can reveal access patterns invisible to generic kernel heuristics~\cite{graefe2006b,qian2024combining,chen2001improving,chen2002fractal,gray1992transaction,ding2007diskseen,ryu2023fast,joo2011fast,lu2024streamsys,lu2026bridging,li2020dadi}. Application directed and cross-layer designs instead expose hints or interfaces that convey such knowledge to the storage stack, allowing applications or runtimes to guide prefetching beyond purely kernel-observed access patterns~\cite{patterson1995informed,yang2002decoupled,vandebogart2009reducing,ge2022hintstor,garg2024crossprefetch}. However, two important gaps still remain in practice. For prediction, workload-specific logic is often tied to particular applications or hint types, limiting reuse across different systems. For execution, prefetch handling is similarly system-specific, duplicating mechanisms and limiting efficient coordination of the page-cache state, in-flight I/O, and shared storage resources.

Together, these two gaps motivate a common OS abstraction that separates workload-aware prediction from prefetch execution. We present \NAME{}, a Linux framework that realizes this abstraction. \NAME{} structures prefetching around two complementary paths: an event path that exports prediction-relevant observations to the user-space runtime, and a request path that imports predicted page ranges into the kernel backend for execution. These paths define two framework-level interfaces: workload events encapsulate the signals from which predictors infer future accesses, while prefetch requests express predictor outputs as target files and bounded page ranges. By standardizing this boundary, \NAME{} lets predictors use workload-specific logic while the operating system retains control over how predictions are validated, scheduled, and executed.



To demonstrate that the same mechanism supports substantially different forms of predictability, we instantiate three policies. A history-based predictor learns recurring sparse file-access sequences during application launch without modifying applications. A SQLite-aware predictor uses B-tree traversal hints to prefetch upcoming child pages during range scans. An LLM-aware predictor uses model-layout metadata to prefetch file ranges needed during model loading. These policies span kernel-observed history, application-provided structural hints, and runtime metadata across diverse workloads. They differ in their event sources, prediction logic, and degree of application involvement, but all emit the same page-range requests and reuse the same bounded kernel backend.

We implement \NAME{} as a Linux prototype using eBPF-based observation for kernel-visible accesses, lightweight user-space channels for optional workload hints, and a shared-memory request interface to a bounded asynchronous kernel backend. Microbenchmarks show that event delivery and user-space dispatch sustain high throughput with low per-event overhead. Across three workload families covering application launch, SQLite range scans, and llama.cpp model loading, \NAME{} reduces application cold-start latency by 5.3--26.3\%, accelerates fragmented SQLite range scans by up to 2.99$\times$ over baseline execution, and reduces LLM model-loading latency by 8.8--47.9\% across three models. We further show that these benefits persist across NVMe SSDs, SATA SSDs, and HDDs despite their substantially different latency and bandwidth characteristics. Together, these results demonstrate that a single, reusable OS abstraction can support diverse workload-aware predictors while delivering end-to-end performance gains.

This paper makes the following contributions:
\begin{itemize}[leftmargin=*]

\item We formulate delegated workload-aware prefetching as an OS abstraction problem and present \NAME{}, separating user-space prediction from bounded, OS-controlled execution in the kernel backend.

\item We instantiate \NAME{} with three substantially different prediction policies based on access history, B-tree traversal structure, and LLM tensor-layout metadata, demonstrating a common request interface across workload-specific predictors.

\item We evaluate \NAME{} across microbenchmarks and end-to-end workloads, showing speedups for application launch, SQLite scans, and LLM loading. We further show policy reuse across storage engines, workload-dependent migration of integrated prefetchers, and benefits across heterogeneous storage devices.

\end{itemize}
\section{Background and Motivation}
\label{background}


\subsection{Readahead Sees Locality, Not Structure}

\begin{figure*}[t]
\centering

\begin{minipage}[t]{0.49\textwidth}
    \centering
    \includegraphics[width=0.49\linewidth]{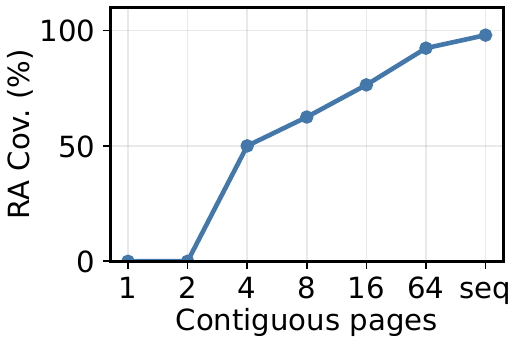}
    \hfill
    \includegraphics[width=0.49\linewidth]{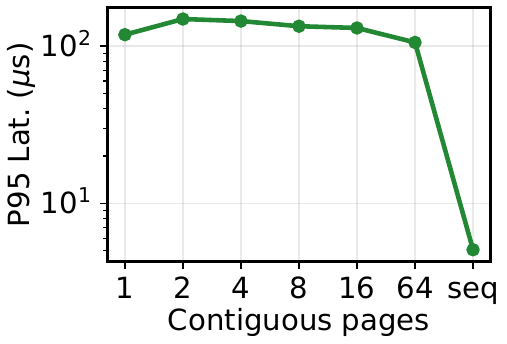}

    \vspace{-0.1cm}
    \caption{Linux readahead exploits offset locality rather than workload structure. Short contiguous runs receive little timely readahead and suffer high demand-read latency, while highly sequential accesses approach the sequential upper bound.}
    \label{fig:bg-ra-cov}
\end{minipage}
\hfill
\begin{minipage}[t]{0.49\textwidth}
    \centering
    \includegraphics[width=0.49\linewidth]{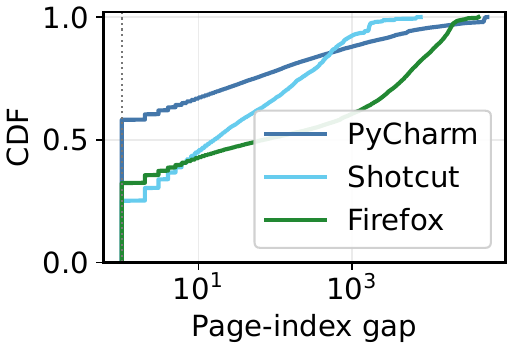}
    \hfill
    \includegraphics[width=0.49\linewidth]{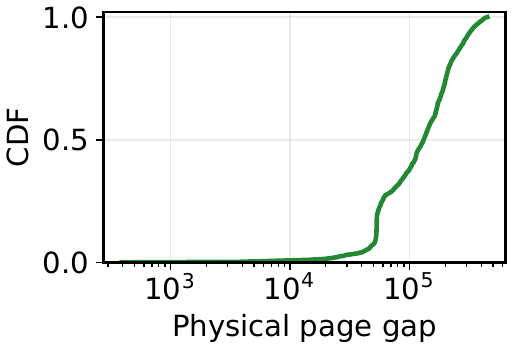}

    \vspace{-0.1cm}
    \caption{Different workloads expose different prediction signals. Application launch contains recurring but sparse per-file page sequences, while SQLite B-tree traversal maps logically adjacent accesses to physically distant pages.}

    \label{fig:bg-predictability}
\end{minipage}

\end{figure*}


Storage prefetching reduces access latency by moving data into memory before demand. In Linux, this role is primarily served by readahead, which speculatively loads nearby file-backed pages into the page cache based on observed offset locality and page-cache state~\cite{wu2007linux,pai2004linux}. This design is deliberately general and works without application changes, making it effective when recent spatial locality is a reliable predictor of future accesses. Variants add more robust heuristics or learning-based predictors~\cite{laga2016lynx,fengguang2008design}, but their inputs remain kernel-visible signals such as recent offset locality, page-cache state, and per-file readahead state rather than workload-specific structure.

Figure~\ref{fig:bg-ra-cov} uses a controlled microbenchmark to isolate readahead's dependence on local sequentiality. Each trial issues 4,096 page-sized reads in total. For run length $r$, the benchmark reads $r$ consecutive pages before jumping to a distant region, while \texttt{seq} issues a fully contiguous stream. Before each demand read, we check whether the page is already resident in the page cache to approximate timely readahead coverage.
Coverage improves with longer runs, but jumping patterns still exhibit high P95 demand-read latency. Each jump starts a new cold region, leaving the first pages of a run on the demand path before readahead catches up. Fully sequential access, in contrast, achieves near-complete coverage and microsecond-scale P95 latency.


Many modern workloads are predictable in ways that do not appear as sustained offset locality. Application launch touches sparse pages across executables, shared libraries, JAR files, and assets, while database range scans traverse B-tree pages according to logical tree topology rather than physical adjacency. These accesses may be structured, repeatable, or known to the application, but they do not form the contiguous runs that Linux readahead is designed to recognize. The limitation is one of visibility: readahead observes file offsets and recent locality, but it does not know that a page is the next child of a B-tree node, or that a sparse startup sequence has recurred across launches. Consequently, workloads that are predictable to an application or runtime may still appear irregular to a storage layer that can observe only file offsets. Improving prefetching for such workloads requires signals beyond generic offset locality.

\subsection{Why One Predictor Cannot Fit All}

The visibility gap described above is not limited to just a single alternative signal. Different workloads often expose predictability through different mechanisms, and these mechanisms require different prediction logic to be turned into useful prefetch requests. Recurrence across executions, logical relationships in data structures, and runtime metadata each provide potentially useful information for some workloads but not others~\cite{ding2007diskseen,graefe2006b,liu2026accelerating}. A predictor designed around one of these signals therefore cannot serve workloads that depend on another.

Figure~\ref{fig:bg-predictability} illustrates this inherent diversity. Application launch exposes recurring but sparse per-file page sequences: the left panel clearly shows that consecutive accesses often jump across non-adjacent page indices, and the gap distribution differs across applications. This structure suits a history-based predictor, because the same startup path can recur across executions even when accesses are not sequential. SQLite range scans expose a different signal: the right panel further shows that pages logically adjacent in a B-tree traversal can be separated by large physical page gaps, so the useful predictor is application knowledge of upcoming tree children rather than recent offset locality.

LLM model loading exposes yet another form: the runtime can derive tensor offsets and sizes from model metadata before those pages are demanded, creating a prefetch window in which I/O overlaps with continued loading. As Table~\ref{tab:bg-llm-io} shows, major faults remain unchanged while filesystem inputs and voluntary context switches drop by 99.6\% and 97.8\%, indicating that faults still occur but no longer wait for storage. This gap shows that most faults no longer wait for storage because the required pages are already in the page cache. The useful signal here is neither sequential locality nor a repeated startup trace, but knowledge of tensor layout and model metadata available to the runtime.

Thus predictability depends heavily on the workload in practice. Historical correlations, structural relationships, and runtime metadata each provide useful information for different workloads, but they require different policies to translate that information into page predictions. The operating system should therefore provide a common abstraction that accommodates diverse prediction inputs while allowing policies to share a common output.

\begin{table}[t]
\centering
\caption{Process-visible I/O behavior during cold LLM loading.}
\label{tab:bg-llm-io}
\begin{tabular}{lrrr}
\toprule
\textbf{Metric} &
\textbf{Baseline} &
\textbf{\NAME{}} &
\textbf{Change} \\
\midrule
Major faults                 & 192       & 190       & -1.0\% \\
Minor faults                 & 1,081,988 & 1,402,301 & +29.6\% \\
context switches   & 19,252    & 428       & -97.8\% \\
Filesystem inputs           & 9,847,984 & 40,360    & -99.6\% \\
\bottomrule
\vspace{-0.5cm}
\end{tabular}
\end{table}

\subsection{Separating Prediction from Execution}

Prefetching involves two distinct functions: predicting which pages will be needed and executing speculative reads to bring them into memory. Traditionally, both functions have been coupled inside kernel readahead: the kernel observes file accesses, predicts future pages, and issues the corresponding I/O. This design works naturally when prediction can be derived from kernel-visible access patterns. Operating systems also expose interfaces such as \texttt{fadvise()}, \texttt{madvise()}, and asynchronous I/O~\cite{kim2016nvmedirect,lee2019asynchronous,shin2014path,nicolas2025investigating,didona2022understanding}, allowing applications to influence or initiate data fetching while the kernel still performs the underlying I/O. However, these interfaces primarily express low-level access hints or file ranges rather than workload structure. Consequently, systems that exploit richer knowledge increasingly move prediction logic outside the kernel. This evolution shows that prediction and prefetch execution need not reside in the same layer.


These two functions are therefore different responsibilities. Applications and user-space runtimes are better positioned to determine which pages are likely to be useful because they understand workload structure and execution context. Much of this information is unavailable to the kernel, while user-space policies can evolve independently of kernel changes. The operating system, in contrast, is better positioned to execute speculative reads because it owns the page cache, observes system-wide I/O state, can suppress redundant requests for cached or in-flight pages, and can bound speculative work to limit cache pollution and bandwidth waste. It can further coordinate prefetch activity across workloads sharing the same storage resources. Reimplementing these execution mechanisms within each application duplicates effort and couples workload-specific predictors to application-specific I/O paths, making them harder to reuse.

This observation suggests a policy--mechanism split~\cite{humphries2021ghost,260149,garg2024crossprefetch,liu2026accelerating}. Prediction policies should evolve in user space, where they can incorporate workload-specific knowledge without kernel modification, while a common kernel mechanism validates predictions against current page-cache state and performs bounded asynchronous prefetch execution. \NAME{} adopts this split, enabling diverse workload-aware predictors to share a common execution substrate without embedding workload-specific policy in the kernel.

\section{Design}

\begin{figure}[!t]
    \centering
    \includegraphics[width=0.98\linewidth]{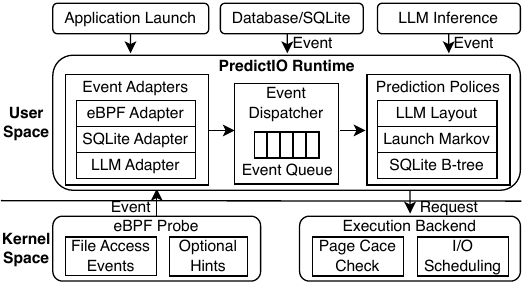}
    \caption{\NAME{} architecture. Kernel observations and application-provided hints feed workload-specific prediction policies in user space. The resulting page-range requests are executed by a shared, bounded, page-cache-aware kernel backend.}
    \label{fig:design-architec}
    \vspace{-0.3cm}
\end{figure}

Figure~\ref{fig:design-architec} presents the architecture of \NAME{}. The design separates workload-aware prediction from prefetch execution through two paths. The event path carries prediction-relevant signals from kernel observations or application-provided hints into the user-space runtime, where event adapters and a dispatcher route them to workload-specific policies that determine what to prefetch. The request path carries the resulting file-page ranges to a shared execution backend. This separation allows prediction policies to vary with workload structure while keeping prefetch execution common and OS-controlled.

\subsection{Design Goals}

\NAME{} is guided by four design goals that make diverse workload-aware predictors practical under a shared, bounded execution mechanism:

\begin{enumerate}[leftmargin=*]
\item \textbf{Heterogeneous prediction signals.}
Support diverse inputs, including access history and other workload-specific signals, while allowing policies to emit a common page-range request format.
\item \textbf{User-space prediction policies.} Keep workload-specific prediction logic in user space, where it can evolve independently of the kernel.
\item \textbf{Kernel-resident execution.} Retain page-cache-aware prefetch execution in the kernel, which can avoid redundant I/O and populate the shared page cache.
\item \textbf{Bounded speculative work.} Limit queued requests, request sizes, and in-flight prefetches to prevent inaccurate predictions from injecting unbounded I/O.
\end{enumerate}

\begin{table}[t]
\centering
\small
\caption{Event families used by the current \NAME{} prototype.}
\label{tab:event-families}
\begin{tabular}{lll}
\toprule
\textbf{Event family} & \textbf{Signal} & \textbf{Example policy} \\
\midrule
File access        & File ID and offset & History-based launch \\
B-tree traversal   & Upcoming child pages & Tree-aware range scan \\
Tensor layout      & Tensor file ranges & Layout-aware loading \\
\bottomrule
\end{tabular}
\end{table}

\subsection{Workload Event Sources}

\NAME{} accepts workload events from two sources: kernel observation for unmodified applications and application-provided hints for workloads that can expose additional prediction signals. Both sources deliver their signals to the runtime through a corresponding adapter. Table~\ref{tab:event-families} summarizes the event families supported by our current prototype.

For unmodified applications, \NAME{} uses kernel hooks to observe page-level file accesses, including mmap-driven accesses not exposed through user-space I/O interfaces. In our prototype, eBPF probes attach to file-backed access paths and deliver events to the runtime through an eBPF ring buffer. These events are received by the eBPF adapter and forwarded into the runtime event path. Each event identifies the process, the accessed file, and the page offset. These events expose recurring access sequences across executions. This path is useful for application launch, where accesses may be sparse but repeat across cold starts. The same event-source interface can be extended with additional kernel hooks when other prediction signals are useful.

An application hint describes workload structure that becomes available before the corresponding pages are demanded and is delivered to the runtime as a workload event. Our prototype defines typed hints for B-tree traversal and tensor layout. The SQLite runtime emits a tree-child hint when a range scan descends through an internal B-tree page, identifying the database, the parent page, the current child position, and the upcoming logical children. Other B-tree based systems could expose the same hint family. The LLM runtime emits tensor-layout hints that describe the file offsets and sizes of model tensors. Because model loading follows a known execution order, these hints allow the policy to predict which file ranges will be needed before the corresponding tensors are demanded. Each hint is delivered to its corresponding adapter in the runtime.

\subsection{User-Space Prediction Policies}

The user-space runtime realizes the event-to-request abstraction between workload-specific signals and kernel prefetch execution. Event adapters translate source-specific observations or hints into typed \emph{workload events}. Each event carries the information needed by its corresponding prediction policy, while hiding how that information was obtained. The dispatcher receives these events and routes them to policies that accept the corresponding event type. This interface allows kernel observations, application hints, and other event sources to share the same prediction pipeline.

A prediction policy consumes workload events and translates them into predicted page accesses. Each policy implements an event handler that may update policy-local state and emit zero or more prefetch requests. The framework does not prescribe how predictions are derived; policies may exploit structural relationships or other workload-specific knowledge. Policies decide only what to prefetch and do not manage page-cache state or issue storage I/O, keeping prediction independent of execution. Section~\ref{sec:policies} describes the concrete policies used in our prototype.

The output of every policy is a \emph{prefetch request}, which identifies a target file and a set of predicted page offsets or ranges. This common request format forms the boundary between prediction and execution. Before submission, \NAME{} sorts and deduplicates sparse page lists and coalesces adjacent pages into ranges where possible. The normalized request is then handed to a backend for execution. Thus, heterogeneous workload events may trigger different prediction logic, but all policies ultimately communicate through the same page-range request interface.

\vspace{-0.2cm}

\subsection{Kernel-Space Execution Backend}

A backend executes the prefetch requests produced by prediction policies and may reside in user space or in the kernel. Our prototype implements a kernel-space backend so that execution can coordinate with page-cache state, in-flight I/O, and competing workloads. This backend provides a shared execution substrate for all policies: prefetch requests arriving from the shared ring are validated, partitioned into bounded work items, and admitted to a bounded queue, which a configurable pool of kernel workers drains asynchronously.

For each work item, the backend checks the current page-cache and in-flight state before issuing I/O. Pages that are already resident or already being fetched are skipped, while the remaining pages are submitted as asynchronous reads to populate the page cache. This filtering allows different policies to share the same execution path without generating redundant storage requests.

Execution is bounded at both admission and I/O stages. Queue capacity limits the amount of pending speculative work, while an in-flight-page limit bounds the amount of prefetch I/O issued to storage. Excess speculative work can therefore be dropped or deferred without blocking the application. If a prediction is late or omitted, normal demand paging remains the fallback path. These mechanisms limit the impact of inaccurate predictions while remaining independent of any particular prediction policy.
\section{Policy Instantiations}
\label{sec:policies}

This section illustrates how the same workload event and prefetch request abstractions can be used to design three representative instantiations: a history-based predictor for application launch, a tree-aware predictor for B-tree range scans, and a layout-aware predictor for LLM loading. For each, we describe the prediction opportunity, event source and integration cost, prediction logic, and reuse of the common execution path.

\subsection{History-Based Application Launch}
\label{sec:policy-launch}

Application launch often exhibits recurring file-access patterns. Before becoming responsive, a desktop application may touch sparse pages across executables, shared libraries, JAR files, and assets. These accesses are difficult for locality-based readahead to anticipate, but their order frequently repeats across cold starts of the same application version.

The policy consumes kernel-observed \texttt{FileAccess} events, each of which identifies the target file and accessed page offset. Unlike the semantic policies described later, the launch policy relies on an offline training stage. Rather than replaying a single recorded trace, it aggregates page-access sequences across multiple launches and constructs a first-order Markov model for each file observed during launch. For each observed page, the model stores a probability distribution over pages that previously followed it in the collected traces. By learning transition probabilities across multiple launches, the policy tolerates small variations in startup behavior while still capturing recurring sparse access patterns.

Figure~\ref{fig:launch-policy} shows the online prediction logic. When an access event arrives, the policy locates the per-file model and begins a bounded random walk from the observed page. At each step, it draws a random threshold and walks the cumulative distribution function of the current page's outgoing transitions until the threshold is reached. The selected page becomes the next state and is appended to the prediction if it has not already been seen. The walk stops after a fixed lookahead depth or when no unseen transition is selected. The resulting sparse page list is sorted, deduplicated, and submitted as a standard \NAME{} prefetch request.

The policy requires no modification to the target application. Both training and inference rely exclusively on kernel-observed file-access events. It therefore provides a compatibility path for existing applications while reusing the same event path, request interface, and bounded kernel backend as the semantic policies described next.

\begin{figure}[t]
\centering
\begin{lstlisting}[style=polyfetchcpp]
void LaunchPolicy::OnEvent(Event e) {
  MarkovModel *m = models.Lookup(e.file);
  Page cur = e.page; Pages pages = {cur};

  // Walk a bounded number of transitions from the current page.
  for (int i = 0; i < lookahead; i++) {
    double r = Random(),cdf = 0.0;// choose one random threshold
    // Walk the transition CDF.
    for (auto [next, prob] : m->Trans(cur)) {
      cdf += prob;           // sample by transition probability
      if (r > cdf) continue;
      if (seen.Contains(next)) break;
      seen.Insert(next); pages.Append(next);
      cur = next;
      break;
    }
  }
  pages.SortAndDedup();         // normalize page-list request
  Submit(Req(e.file, pages));   // common request interface
}
\end{lstlisting}
\caption{History-based application-launch policy. The policy samples a bounded
walk over a per-file Markov model, suppresses repeated pages, and submits the
predicted page list through the common \NAME{} request interface.}
\label{fig:launch-policy}
\vspace{-0.3cm}
\end{figure}

\subsection{B-Tree Traversal Range Scans}

B-tree range scans are often logically predictable even when they lack spatial locality. During an ordered traversal, the engine visits child pages according to key order and tree structure. These pages need not be contiguous in the backing file. Page allocation, splits, and updates may place logically adjacent nodes at distant offsets. Linux readahead therefore observes an irregular access stream even though the engine already knows likely future child pages. Storage engines can expose this information through a lightweight \texttt{TreeChildEnter} hint that identifies upcoming child pages during traversal. Our prototype emits this hint in SQLite through an adapter in the B-tree cursor path.

Figure~\ref{fig:sqlite-policy} shows the prediction logic. For each \texttt{TreeChildEnter} event, the policy selects at most \texttt{window} upcoming child pages from the hint. The event exposes these children as file page offsets or ranges, so the policy does not depend on database-specific page numbering or layout details. Successive traversal events may expose overlapping future children, so the policy keeps a short-lived record of recently submitted pages and suppresses repeated predictions for the same file within a configurable time-to-live window. This filter reduces redundant requests before they reach the kernel, while the kernel backend still performs authoritative page-cache and in-flight filtering. The remaining pages are sorted, deduplicated, and submitted as a standard \NAME{} prefetch request.

The policy itself is independent of any particular database engine. Its only requirement is an event that exposes future child pages during B-tree traversal. In our prototype, SQLite provides this information through a lightweight adapter in the cursor path, but the same policy can be reused by other B-tree-based storage systems that expose an equivalent hint. The adapter forwards structural information already available during traversal and does not issue I/O or manage page-cache state. Once emitted, these events reuse the same event path, request interface, and bounded kernel execution mechanism as all other \NAME{} policies.

\begin{figure}[t]
\centering
\begin{lstlisting}[style=polyfetchcpp]
void BTreePolicy::OnEvent(Event e) {
  File db = e.database;
  Pages pages;

  // Prefetch a bounded window of upcoming B-tree children.
  for (Page child : e.future_children.Take(window)) {
    // Convert each logical child page to kernel pages.
    Pages child_pages = ToKernelPages(child, e.page_size);

    for (Page page : child_pages) {
      // Suppress recently submitted pages.
      if (recent.Contains(db, page, ttl)) continue;
      recent.Insert(db, page);
      pages.Append(page);
    }
  }

  pages.SortAndDedup();         // normalize page-list request
  Submit(Req(db, pages));       // common request interface
}
\end{lstlisting}
\caption{B-tree traversal policy instantiated for SQLite. The policy maps a bounded set of upcoming child pages to kernel pages and suppresses recent duplicates before submitting a standard \NAME{} request.}
\label{fig:sqlite-policy}
\vspace{-0.3cm}
\end{figure}

\subsection{LLM Tensor Access}

Large-model inference exposes a different form of predictability. Model files contain tensors associated with different layers, and runtimes typically know both the tensor layout within model files and the order in which layers will execute. This knowledge creates a prefetch window in which pages required by future layers can be fetched while earlier layers continue executing. The runtime can therefore identify future accesses before they occur. It exposes the corresponding tensor boundaries and execution phases as prediction signals that are otherwise invisible to the storage stack. Even when tensors are distributed across multiple regions or shard files, the runtime already knows where they reside. Prefetching can then move storage activity off the critical execution path by overlapping I/O with computation.

The LLM runtime emits a typed \texttt{TensorLayout} hint for each tensor discovered during model metadata parsing. This hint is forwarded by the LLM adapter into the runtime and dispatched to the LLM policy as a \texttt{TensorLayout} event. It identifies the model, tensor name, backing file, byte offset, tensor size, and page size. The adapter only forwards layout information. It does not issue I/O or manage page-cache state.
Figure~\ref{fig:llm-policy} shows the layout-aware policy. For each \texttt{TensorLayout} event, the policy first constructs a tensor key from the backing file, tensor name, offset, and size. Duplicate layout events for the same tensor extent are ignored. The policy then converts the tensor byte extent into inclusive file-page bounds, splits long tensors into bounded page-range entries, and submits a standard \NAME{} request for the backing file. This directly maps model metadata to the page-range request interface used by the kernel backend.

\begin{figure}[t]
\centering
\begin{lstlisting}[style=polyfetchcpp]
void LLMPolicy::OnEvent(Event e) {
  TensorLayout t = e.tensor_layout;
  File file = t.file_path ? t.file_path : t.model_path;
  //Suppress duplicate layout events for the same tensor extent.
  TensorKey key = {file, t.tensor_name, t.offset, t.size};
  if (seen.Contains(key)) return;
  seen.Insert(key);
  // Convert tensor byte extent to inclusive file-page bounds.
  uint64_t end = t.offset + t.size - 1;
  Page first = t.offset / t.page_size;
  Page last  = end / t.page_size;
  Ranges ranges;
  for (Page page = first; page <= last; ) {
    // Split long tensors into bounded range entries.
    int len = Min(MAX_RANGE_LEN, last - page + 1);
    ranges.Append(Range(page, len));
    page += len;
  }
  Submit(Req(file, ranges));   // common request interface
}
\end{lstlisting}
\caption{LLM layout-aware policy. The policy suppresses duplicate tensor layouts, maps tensor byte extents to bounded file-page ranges, and submits a \NAME{} request.}
\label{fig:llm-policy}
\vspace{-0.2cm}
\end{figure}

The policy requires only a lightweight runtime adapter that forwards tensor-layout metadata already parsed during model loading. It does not require changes to the model format or the kernel execution path. Compared with the launch and SQLite policies, this instantiation uses static layout metadata rather than learned history or traversal hints, but it reuses the same event path, request interface, and bounded kernel backend.

\section{Implementation}
\label{sec:implementation}

We implement \NAME{} on Linux as a Rust user-space runtime, a set of eBPF observation programs comprising 269 lines of hand-written code, and a loadable kernel module comprising 2,314 lines of C. The current user-space runtime contains approximately 6,900 non-comment source lines, excluding standalone benchmark binaries and generated bindings. Besides SQLite and llama.cpp, we integrate LMDB~\cite{lmdb} and RocksDB~\cite{rocksdb} to evaluate policy reuse across storage systems. Table~\ref{tab:adapter-loc} summarizes their integration points and added code size. All counts exclude evaluation scripts, generated code, and uncompiled sources.

\subsection{Event Observation and Application Adapters}
\label{sec:implementation-events}

\NAME{} obtains prediction signals from two sources. For unmodified applications, eBPF probes observe kernel-visible file accesses without modifying the kernel or the application. For application launch, a kprobe on \texttt{do\_sync\_mmap\_readahead()} captures file-backed mmap accesses during program and shared-library loading, recording the process ID, file identity, page offset, and kernel timestamp. An allowlist and blocklist filter events to relevant files. Accepted events are serialized as fixed-size \texttt{DentryEvent}s into a BPF ring buffer, dropping events when the buffer is full rather than delaying the application. The user-space adapter decodes each record into a typed \texttt{FileAccess} event and forwards it to the runtime.

For instrumented applications, lightweight adapters expose prediction hints that are invisible to the kernel. Our prototype includes adapters for SQLite, LMDB, RocksDB, and llama.cpp. The database adapters report typed tree-traversal events, while the llama.cpp adapter reports tensor-layout and layer-start events. These hints are serialized and sent through Unix-domain datagram sockets; the runtime decodes them and forwards them to the common event queue. The llama.cpp adapter requires more integration code because its cues span both model loading and inference; the database adapters, in contrast, only report B-tree traversal events.

\begin{table}[t]
  \centering
  \caption{Size of application-specific adapters for \NAME{}.}
  \label{tab:adapter-loc}
  \small
  \begin{tabular}{l l r}
    \toprule
    \textbf{Adapter} & \textbf{Integration point} & \textbf{Added LOC} \\
    \midrule
    SQLite   & B-tree traversal events              & 26  \\
    LMDB     & B-tree traversal events              & 98  \\
    RocksDB  & Block prefetcher integration         & 69  \\
    llama.cpp & Tensor-layout and layer-start events & 266 \\
    \midrule
    \textbf{Total} & & \textbf{459} \\
    \bottomrule
  \end{tabular}
\end{table}

\subsection{Runtime and Policy Architecture}
\label{sec:implementation-runtime}

The \NAME{} runtime decouples event observation, policy execution, and prefetch submission through typed interfaces. Implemented in Rust, it exposes a common event representation and defines policies through a trait with event-matching and event-processing methods. Multiple policies can coexist in one runtime, and each incoming event is dispatched to the policies that accept it. A policy produces zero or more prefetch requests, which are forwarded to a pluggable backend.
The kernel backend converts page predictions into sorted, deduplicated, and coalesced ranges, splits large requests into bounded chunks, and writes them to a shared-memory ring exposed by the kernel module. It then uses a lightweight \texttt{ioctl()} notification to wake the kernel consumer, which validates and forwards the requests to the asynchronous execution queue. When the ring is full, the backend drops the affected chunk instead of blocking the application.

\subsection{Asynchronous Prefetch Execution}
\label{sec:implementation-execution}

The kernel backend converts admitted page-range predictions into asynchronous reads that populate the page cache without interpreting workload semantics. Incoming predictions are split into bounded work items and placed in a FIFO queue serviced by kernel worker threads. Queue occupancy and in-flight prefetch pages are both limited, so excess work is dropped or skipped rather than allowed to inject unbounded I/O.
Before issuing a read, the backend skips pages that are already cached, beyond end-of-file, or located in sparse holes. For remaining pages, it inserts page-cache entries and submits asynchronous \texttt{REQ\_RAHEAD} BIOs. Completion callbacks mark successfully read pages as up to date and release resources.
By default, \NAME{} submits pages independently to preserve sparse access patterns. An optional batching mode coalesces physically consecutive pages into multi-page BIOs. The backend targets block-backed filesystems with \texttt{bmap()} support and records counters for submission, completion, cache hits, throttling, failures, queue occupancy, and latency.
\section{Evaluation}
\label{evaluation}

\noindent
\textbf{Evaluation Goals.}
We evaluate \NAME{} to answer the following questions:
\begin{itemize}[leftmargin=*]
    \item \textbf{Q1: Control-Path Overhead.}
    What overhead does user-space policy delegation introduce along \NAME{}'s control path?

    \item \textbf{Q2: Application Performance.}
    Can applications improve their performance by using \NAME{} across application launch, database traversal, and LLM workloads?

    \item \textbf{Q3: Policy Reuse and Migration.}
    Can an existing policy be reused across different systems, and can an application-integrated policy be migrated into \NAME{} while retaining its benefit?

    \item \textbf{Q4: Storage-Device Sensitivity.}
    How does \NAME{}'s benefit change across storage devices with different latency and bandwidth characteristics?
\end{itemize}

\paragraph{Testbed.}
We conduct our experiments on a Dell Precision 3650 workstation with a 16-core Intel Core i9-11900K processor, 32\,GB of DDR4 memory, and a 256\,GB NVMe SSD. The machine runs Linux kernel 6.6.138 with an Ext4 filesystem. \NAME{} is compiled in release mode using the default configuration in Section~\ref{sec:implementation}.

\paragraph{Methodology.} For cold-cache experiments, we run \texttt{sync} and clear the Linux page cache before each trial by writing \texttt{3} to \texttt{/proc/sys/vm/drop\_caches}. We report the median across three to five independent trials, depending on the experiment. Framework microbenchmarks isolate control-path costs, while end-to-end experiments exercise the complete \NAME{} pipeline.

\paragraph{Workloads.}
We evaluate \NAME{} using workloads that expose different forms of application-level predictability. Our end-to-end benchmarks cover recurring sparse accesses during application launch, structurally related pages during SQLite B-tree traversal, and runtime-visible tensor-layout metadata during LLM model loading. To evaluate policy generality, we additionally use LMDB to study reuse of the same tree-aware policy across storage engines, and RocksDB to study migration of an application-integrated readahead policy into \NAME{}.

\begin{figure}[t]
	\centering
	\begin{subfigure}[t]{0.49\linewidth}
		\begin{minipage}[t]{1\linewidth}
			\includegraphics[width=1\linewidth]{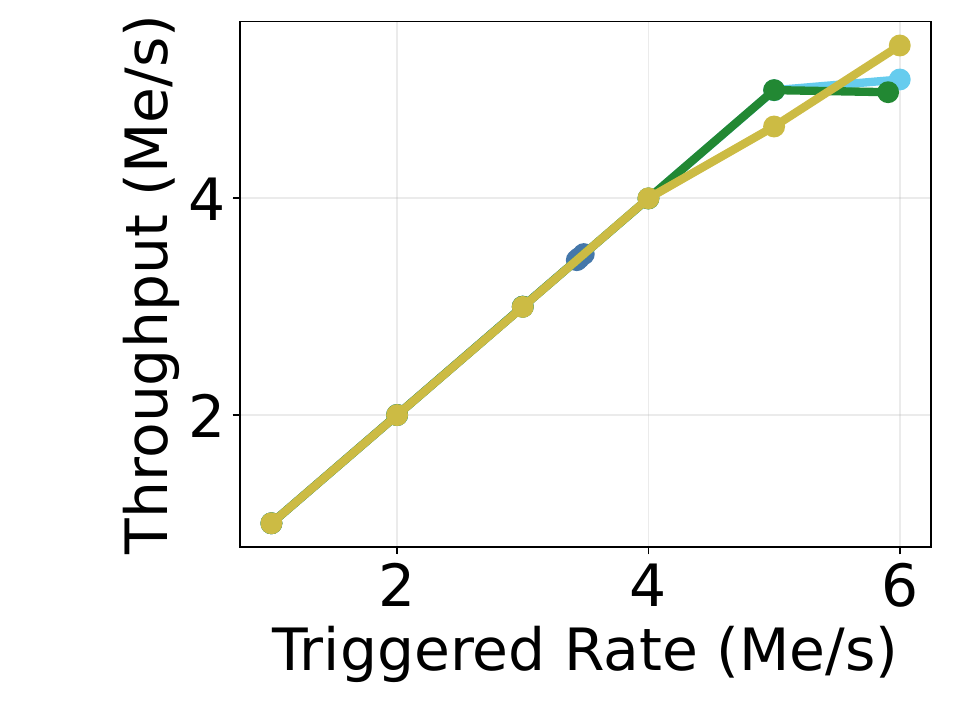}
		\end{minipage}
    \end{subfigure}
    \begin{subfigure}[t]{0.49\linewidth}
		\begin{minipage}[t]{1\linewidth}
			\includegraphics[width=1\linewidth]{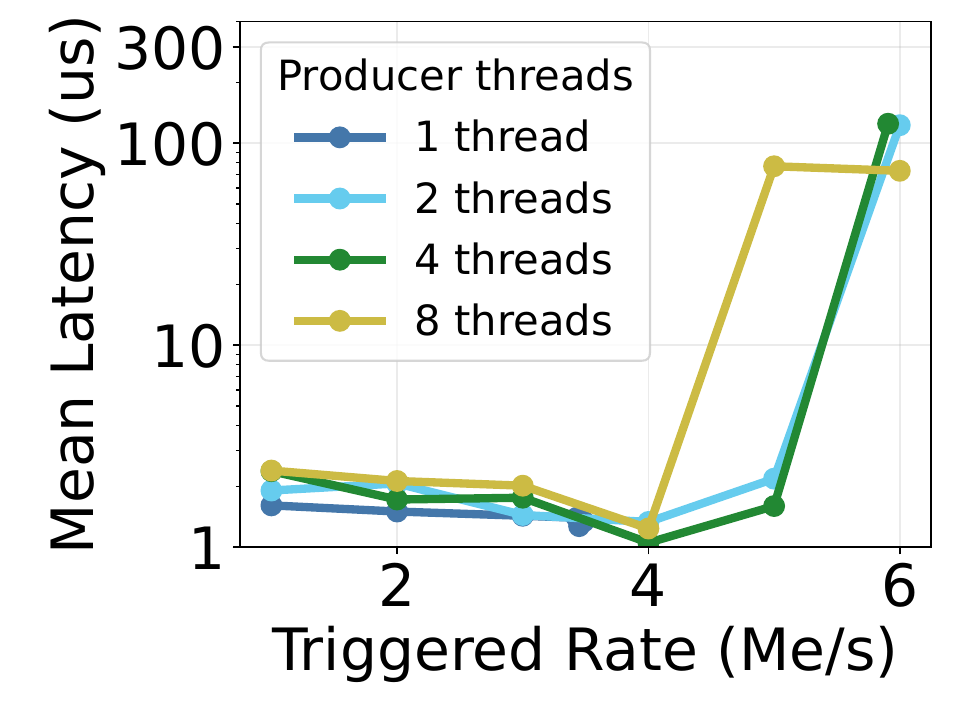}
		\end{minipage}
	\end{subfigure}
	\vspace{-0.4cm}
    \caption{eBPF event-source performance.}
    \vspace{-0.3cm}
    \label{fig:eval-ebpf-source-sweep}    
\end{figure}

\begin{figure}[t]
	\centering
	\begin{subfigure}[t]{0.49\linewidth}
		\begin{minipage}[t]{1\linewidth}
			\includegraphics[width=1\linewidth]{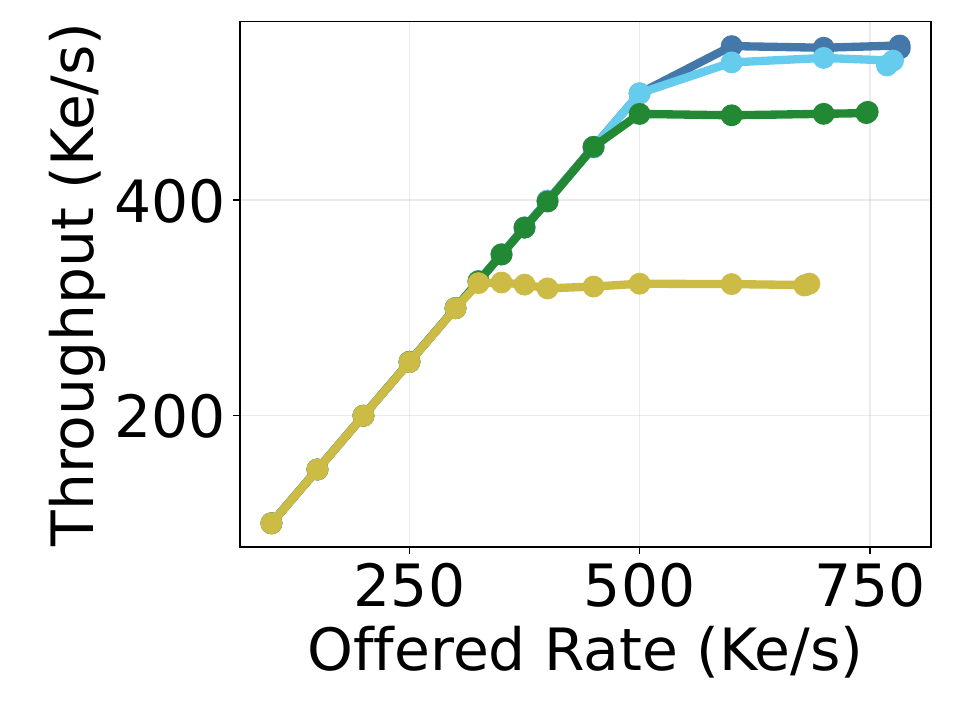}
		\end{minipage}
    \end{subfigure}
    \begin{subfigure}[t]{0.49\linewidth}
		\begin{minipage}[t]{1\linewidth}
			\includegraphics[width=1\linewidth]{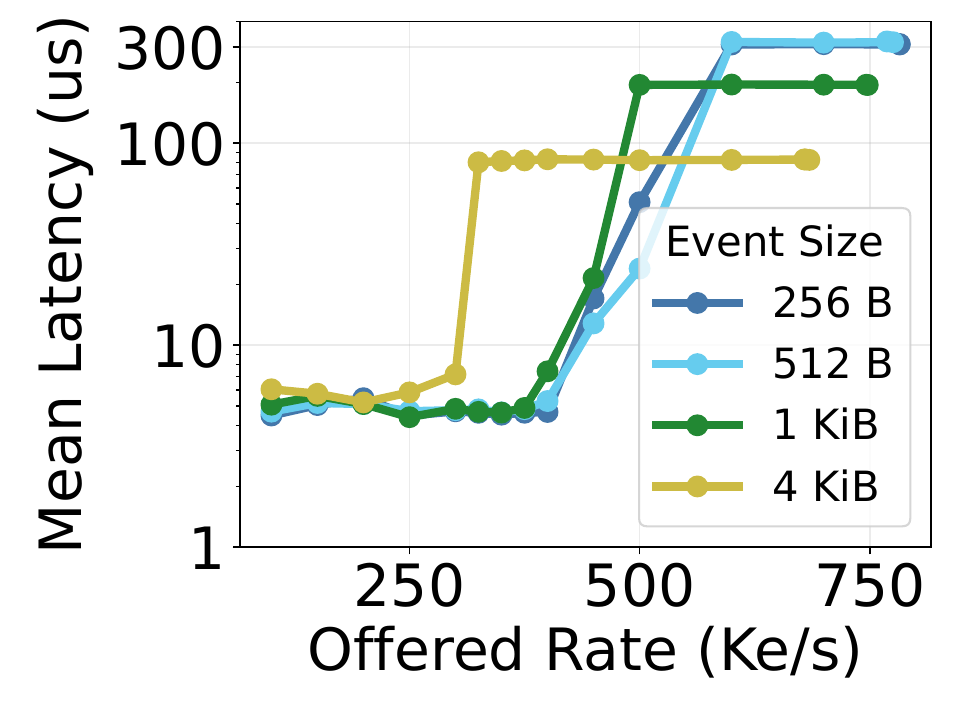}
		\end{minipage}
	\end{subfigure}
	\vspace{-0.4cm}
    \caption{Application event-source performance.}
    \vspace{-0.3cm}
    \label{fig:eval-event-source-sweep}    
\end{figure}

\subsection{Control-Path Overhead}
\label{sec:eval-control-path}

We measure the control-path overhead of \NAME{}: delivering workload events, dispatching them through the runtime, and submitting prefetch requests. Unless otherwise stated, these microbenchmarks use synthetic inputs and disable prefetch execution, allowing us to isolate individual components of \NAME{}'s control path. We report throughput and mean service time while increasing the offered event rate until the source or queue saturates.

\textbf{\textit{Event sources.}} We evaluate the capacity and overload behavior of \NAME{}'s two event-delivery paths: kernel-visible eBPF events and application-level hints. Figure~\ref{fig:eval-ebpf-source-sweep} measures the eBPF path by varying the event rate while using a \texttt{noop} policy and backend. Throughput follows the triggered rate up to approximately 5\,M events/s, with latency remaining low below this point. Figure~\ref{fig:eval-event-source-sweep} measures the application-hint path by sending fixed-size events through Unix-domain datagram sockets. Throughput reaches approximately 300--500\,K events/s, with 4\,KiB events saturating near 320\,K events/s, and latency stays low until saturation, then rises sharply as the socket or source queue overloads. Across both paths, saturation limits throughput and increases event drops rather than blocking the application. Even under a conservative 32-page prefetch window per event, both paths provide a prefetch-data bound far above the storage bandwidth of the devices we evaluate, and therefore sufficient capacity for the workloads in this paper.

\begin{figure}[t]
	\centering
	\begin{subfigure}[t]{0.49\linewidth}
		\begin{minipage}[t]{1\linewidth}
			\includegraphics[width=1\linewidth]{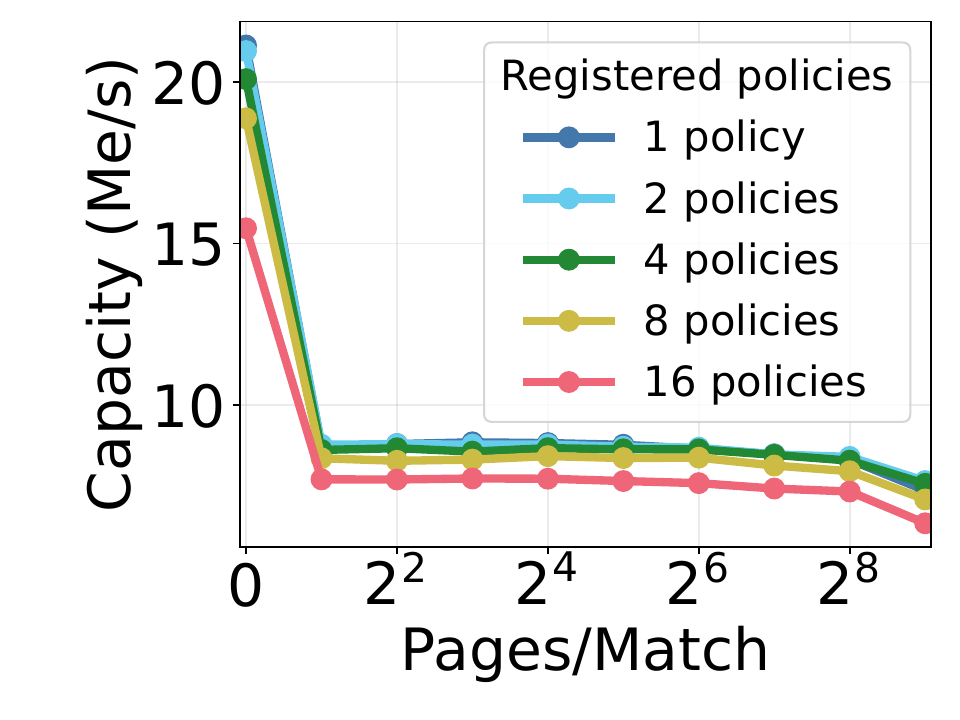}
		\end{minipage}
    \end{subfigure}
    \begin{subfigure}[t]{0.49\linewidth}
		\begin{minipage}[t]{1\linewidth}
			\includegraphics[width=1\linewidth]{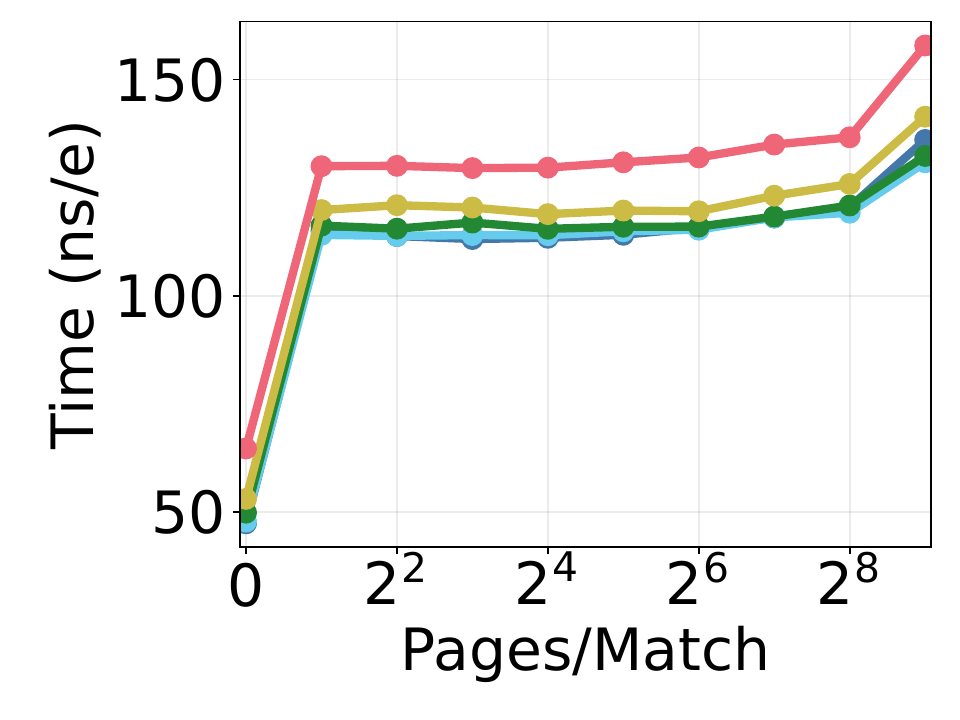}
		\end{minipage}
	\end{subfigure}
	\vspace{-0.4cm}
    \caption{Runtime dispatch performance.}
    \vspace{-0.3cm}
    \label{fig:eval-runtime-performance}    
\end{figure}

\textbf{\textit{Runtime dispatch.}} Figure~\ref{fig:eval-runtime-performance} measures the common runtime with synthetic events and a \texttt{noop} backend. With no policy producing a request, the runtime processes 15.5--21.1\,M events/s across the tested numbers of registered policies. When matching policies emit page predictions, throughput remains between 6.3 and 8.8\,M events/s, with mean service time below 160\,ns per event. The decrease reflects policy matching, request construction, and backend invocation rather than event-source processing. The relatively small change across policy counts indicates that the runtime's trait-based dispatch remains lightweight, while the bounded queue and nonblocking submission path prevent request bursts from directly delaying event observation.

\textbf{\textit{Control-path breakdown.}} We separate user-space processing from kernel-side request processing across the same three workloads. User-space cost is amortized per input event over the entire path, while kernel-side cost measures shared request-ring processing, excluding all I/O-related operations such as file opening, page lookup, and BIO submission.
Table~\ref{tab:control-path-breakdown} reports the median of three independent trials. User-space overhead ranges from roughly 10\,$\mu$s/event for application launch to 0.4--0.45\,ms/event for SQLite and LLM loading, reflecting differences in event representation and policy processing. Kernel-side cost is only tens of ns/page for application launch and SQLite and below 1\,ns/page for LLM loading. The especially small LLM value comes from amortization: its tensor-layout events describe more than 1.2 million predicted pages, so the shared-ring processing time is spread over a much larger page volume. Overall, user-space overhead depends on workload semantics, while kernel request processing remains small when amortized over the pages represented by each workload.

\subsection{End-to-End Performance}
\label{sec:eval-end-to-end}

We next evaluate whether \NAME{}'s delegated prefetching model improves end-to-end application workloads. Unlike the preceding control-path experiments, these measurements run the complete pipeline: workload event generation, user-space dispatch, policy execution, request submission through the shared ring, and bounded prefetch execution in the kernel backend. We compare three modes: \emph{Baseline}, which runs the workload without \NAME{}; \emph{Noop}, which enables the \NAME{} event source and runtime but performs no prefetch I/O; and \emph{\NAME{}}, which runs the workload-specific policy and executes its requests through the kernel backend. Unless otherwise stated, figures report median latency over three cold-cache runs, normalized to Baseline.

\begin{figure}[!t]
    \centering
    \includegraphics[width=0.98\columnwidth]{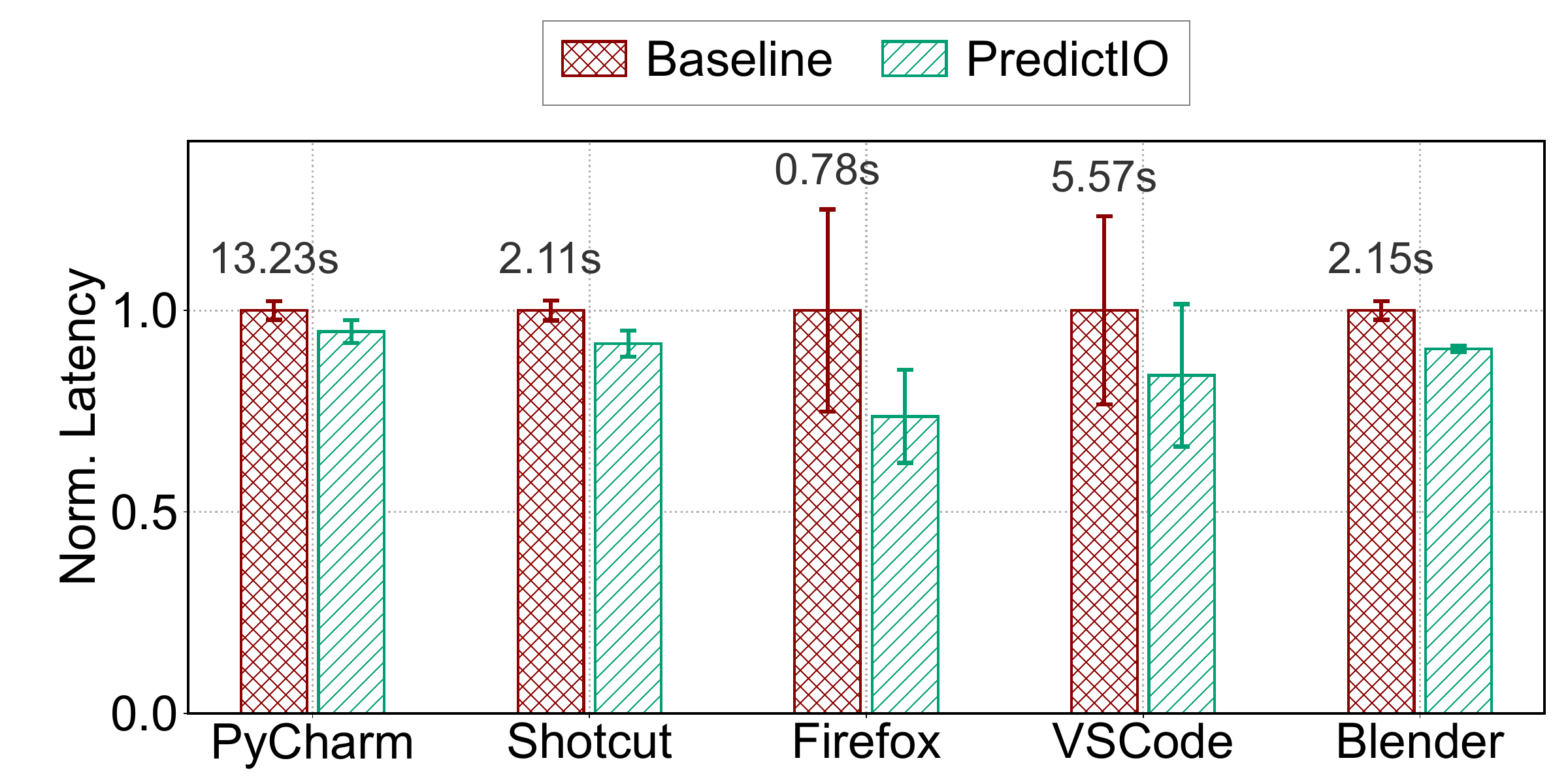}
    \vspace{-0.3cm}
    \captionof{figure}{Application cold-start latency for unmodified desktop applications. Bars are normalized to Baseline; lower is better. Numbers above groups show Baseline mean startup latency.}
    \label{fig:end-to-end-app-start}
    \vspace{-0.3cm}
\end{figure}

\paragraph{Application cold start.} Figure~\ref{fig:end-to-end-app-start} evaluates \NAME{} on five unmodified desktop applications using the history-based launch policy from Section~\ref{sec:policy-launch}. To measure startup latency, we instrument the kernel file-read and page-fault paths, including \texttt{generic\_file\_read\_iter()} and \texttt{do\_fault()}. We compute the interval from the first startup I/O timestamp to when the application UI becomes visible and file-access activity ceases for a sufficiently long quiet period. In these runs, \NAME{} observes kernel-visible file accesses through eBPF and uses per-file Markov models trained from prior launches to prefetch recurring sparse page sequences. Baseline runs the same applications without \NAME{}. Bars report mean normalized latency, and error bars show one standard deviation normalized to the Baseline mean.

\begin{table}[t]
  \centering
  \caption{Control-path overhead across workloads. Counts report input events and predicted pages.}
  \label{tab:control-path-breakdown}
  \small
  \setlength{\tabcolsep}{4pt}
  \begin{tabular}{c|cc|cc}
    \toprule
    \multirow{2}{*}{\textbf{Workload}} &
    \multicolumn{2}{c|}{\textbf{User space}} &
    \multicolumn{2}{c}{\textbf{Kernel backend}} \\
    &
    \textbf{Events} & \textbf{(ns/event)} &
    \textbf{Pages} & \textbf{(ns/page)} \\
    \midrule
    App start   & 1,699  & 9,770   & 13,185    & 22.5   \\
    LLM load    & 435    & 449,570 & 1,226,431 & 0.0546 \\
    SQLite scan & 56,864 & 387,160 & 451,005   & 37.4   \\
    \bottomrule
  \end{tabular}
\end{table}

\NAME{} improves cold-start latency for all five applications, reducing mean latency by 5.3--26.3\% (geometric mean 13.4\%) with the same launch policy, event path, and request interface, and no application-specific tuning. The benefit scales with the share of startup time attributable to sparse, recurring file I/O. Firefox and VSCode benefit most: Firefox concentrates startup I/O in a few large files, and VSCode loads many recurring components, so predictions eliminate a larger fraction of storage stalls. Shotcut touches many smaller files, where metadata and per-file overheads dominate startup time, reducing the gain from data-page prefetching alone. PyCharm sees the smallest reduction, as more of its startup is spent in computation and runtime initialization that prefetching cannot hide. These differences reflect workload structure rather than the framework.

\begin{figure}[t]
	\centering
	\begin{subfigure}[t]{0.49\linewidth}
		\begin{minipage}[t]{1\linewidth}
			\includegraphics[width=1\linewidth]{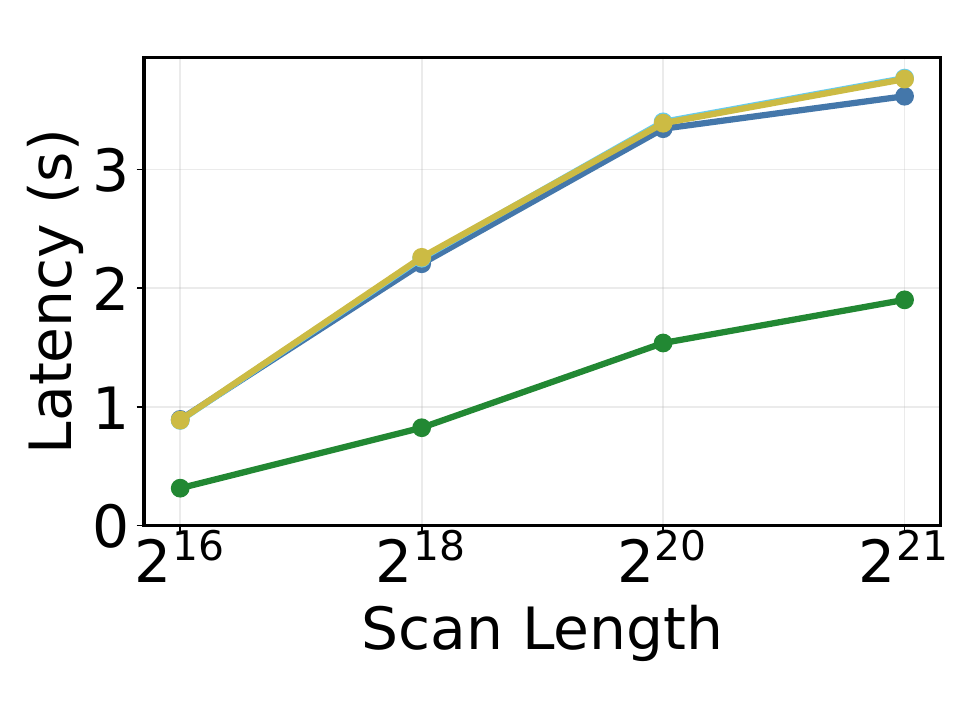}
		\end{minipage}
    \end{subfigure}
    \begin{subfigure}[t]{0.49\linewidth}
		\begin{minipage}[t]{1\linewidth}
			\includegraphics[width=1\linewidth]{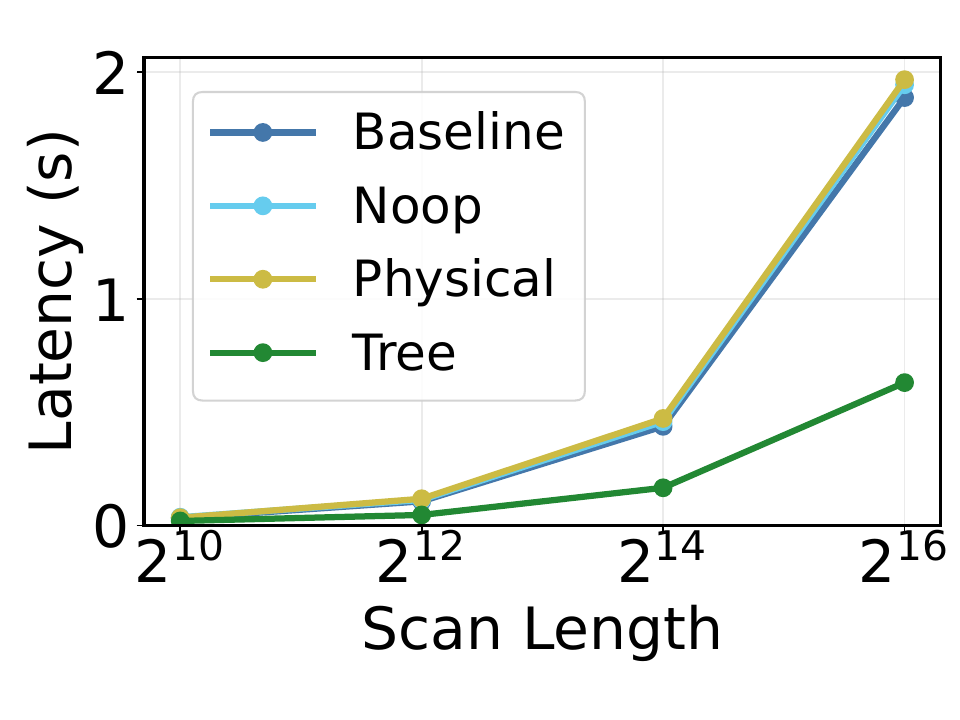}
		\end{minipage}
	\end{subfigure}
	\vspace{-0.4cm}
    \caption{SQLite range-scan latency. Left: controlled fragmented \texttt{kv} scans. Right: YCSB-style scans over a fragmented table. Physical prefetches adjacent pages, while Tree prefetches future B-tree child pages.}
    \vspace{-0.3cm}
    \label{fig:end-to-end-sqlite}    
\end{figure}

\paragraph{SQLite range scans.} Figure~\ref{fig:end-to-end-sqlite} evaluates SQLite range scans with \texttt{mmap} disabled and a small page cache. Both panels compare four modes: Baseline runs SQLite without \NAME{}; the other three run through \NAME{}, differing in prediction policy and request handling. Noop enables the SQLite hint path and tree-children policy but discards requests before the kernel, Physical uses a physical-neighbor policy to prefetch pages adjacent to the current SQLite page, and Tree uses the tree-children policy to prefetch future B-tree child pages exposed by SQLite's cursor.

The left panel uses a controlled fragmented \texttt{kv(id, payload)} database to test whether logical B-tree hints improve range scans. Each query scans a contiguous \texttt{id} interval of $2^{16}$ to $2^{21}$ records, reading every matching record's payload. Noop stays near Baseline, showing limited overhead from the hint path and policy processing, while Tree achieves lower latency and Physical stays near or slightly above Baseline. This shows that application-level B-tree hints can be translated into useful prefetch requests for ordered scans.

The right panel tests whether this behavior persists beyond the controlled fixture on a YCSB-style SQLite database. YCSB's randomized insertion misaligns logical key order from physical page order, the fragmentation \NAME{} targets. The database has one million 1\,KiB records in randomized key order. Each query scans a contiguous \texttt{ycsb\_key} interval of $2^{10}$ to $2^{16}$ records and returns one aggregate value. Tree reduces latency by 38.3--66.6\%, reaching 2.99$\times$ at the smallest scan, while Physical stays near or slightly above Baseline and Noop near Baseline. These results show that the benefit persists on a larger randomized key-value workload, and that \NAME{} can carry SQLite's B-tree hints through its runtime and backend to improve scans.

\begin{figure}[t]
	\centering
	\begin{subfigure}[t]{0.49\linewidth}
		\begin{minipage}[t]{1\linewidth}
			\includegraphics[width=1\linewidth]{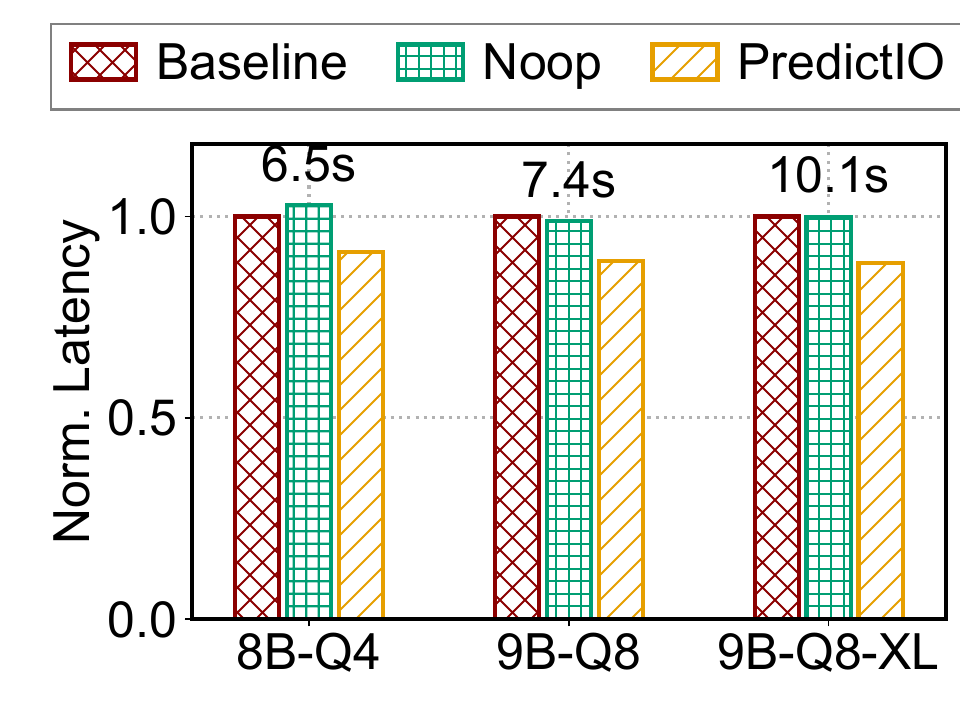}
		\end{minipage}
    \end{subfigure}
    \begin{subfigure}[t]{0.49\linewidth}
		\begin{minipage}[t]{1\linewidth}
			\includegraphics[width=1\linewidth]{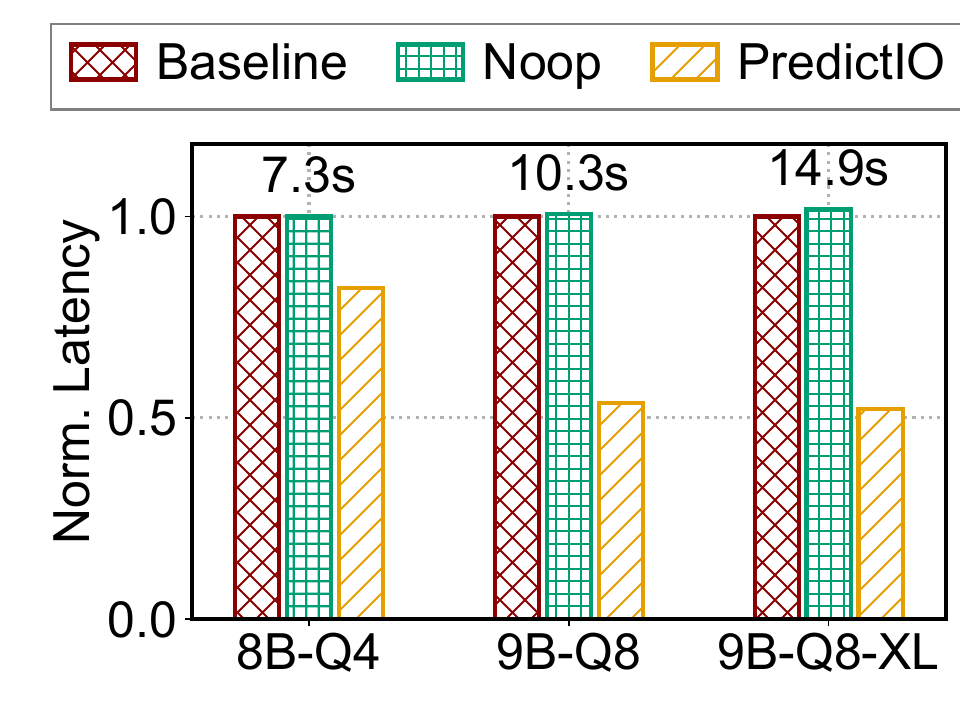}
		\end{minipage}
	\end{subfigure}
	\vspace{-0.3cm}
    \caption{LLM cold-start latency with \texttt{mmap} enabled (left) and disabled (right). Bars are normalized to Baseline; lower is better. Numbers above each group report the Noop latency in seconds.}
    \label{fig:end-to-end-llm}
    \vspace{-0.5cm}
\end{figure}

\paragraph{LLM cold start.} We evaluate three GGUF models for cold start: \texttt{Qwen3-8B-Q4\_K\_M} (4.7\,GiB), \texttt{Qwen3.5-9B-Q8\_0} (8.9\,GiB), and \texttt{Qwen3.5-9B-UD-Q8\_K\_XL} (12.1\,GiB), as shown in Figure~\ref{fig:end-to-end-llm}. We consider two loading modes: default \texttt{mmap}, where model data is fetched on demand through page faults, and an explicit no-\texttt{mmap} path, where tensor contents are explicitly read during loading. The two modes expose different underlying access paths to the kernel, so evaluating both tests whether the same layout-aware policy remains effective across different runtime I/O mechanisms. In both cases, \NAME{} consumes tensor-layout events and prefetches the corresponding model-file ranges through the kernel backend, since GGUF metadata exposes tensor offsets before the runtime consumes tensor contents.

With \texttt{mmap}, \NAME{} reduces latency by 8.8--11.5\% across the three models. The improvement is modest because \texttt{mmap} defers model-data reads to demand faults, leaving less work that layout prefetching can hide. Even so, \NAME{} consistently improves over both Baseline and Noop. The no-\texttt{mmap} path exposes a larger prefetching opportunity: \NAME{} reduces latency by 17.7\% for Q4 and by 46.2--47.9\% for the two Q8 models, while Noop remains within 3\% of Baseline. Thus, within each fixed loading mechanism, the same layout-aware policy improves cold-start performance across different model configurations, while the comparison between the two panels shows that the benefit persists across different llama.cpp loading paths.

\subsection{Policy Reuse and Migration}
\label{sec:eval-policy-reuse}

We evaluate two properties of \NAME{}: policy reuse across applications without per-system rewriting, and migration of an application's own prefetching policy into \NAME{} without losing its behavior or performance benefit.

\paragraph{Policy reuse.}
We first evaluate whether the same policy for a given workload can be reused across different storage systems.
We instantiate the tree-aware policy in SQLite and LMDB, two independently implemented storage engines with different internal implementations and interfaces. The prediction policy itself is unchanged; only a lightweight adapter is used to expose the corresponding traversal events. 
Before each trial, we clear the page cache and normalize latency to the corresponding Baseline.

\begin{figure}[!t]
    \centering
    \includegraphics[width=0.98\columnwidth]{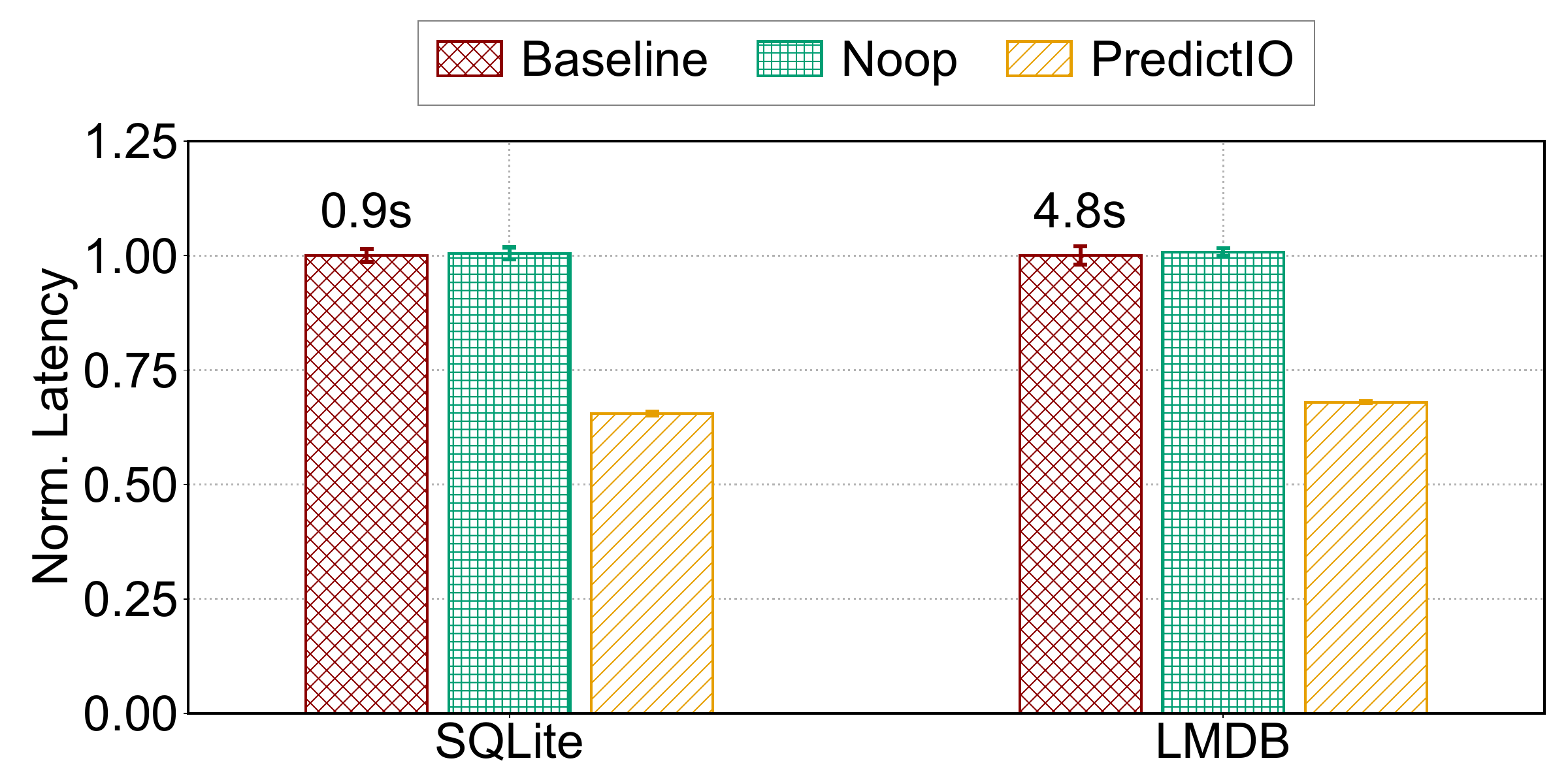}
    \vspace{-0.3cm}
    \caption{Policy reuse across SQLite and LMDB.}
    \label{fig:policy-reuse}
    \vspace{-0.3cm}
\end{figure}

Figure~\ref{fig:policy-reuse} shows that the same policy benefits both systems. \NAME{} reduces latency by 34.3\% for SQLite and 32.0\% for LMDB. The similar gains arise because the structure the policy exploits, namely pages logically adjacent in a B-tree traversal but physically scattered in the backing file, is common to both engines, despite differing page allocation, split behavior, and on-disk layout. Since the prediction logic operates on the traversal structure rather than on engine-specific details, it identifies the same prefetching opportunities in both cases. The residual difference reflects only the two engines' distinct I/O paths and adapter integration, as both submit requests through the same interface and rely on the same kernel backend. These results show that \NAME{}'s event abstraction decouples prediction logic from the underlying storage system: once an engine exposes traversal events through the framework, the same policy can be reused with negligible overhead.

\begin{figure}[!t]
    \centering
    \includegraphics[width=0.98\columnwidth]{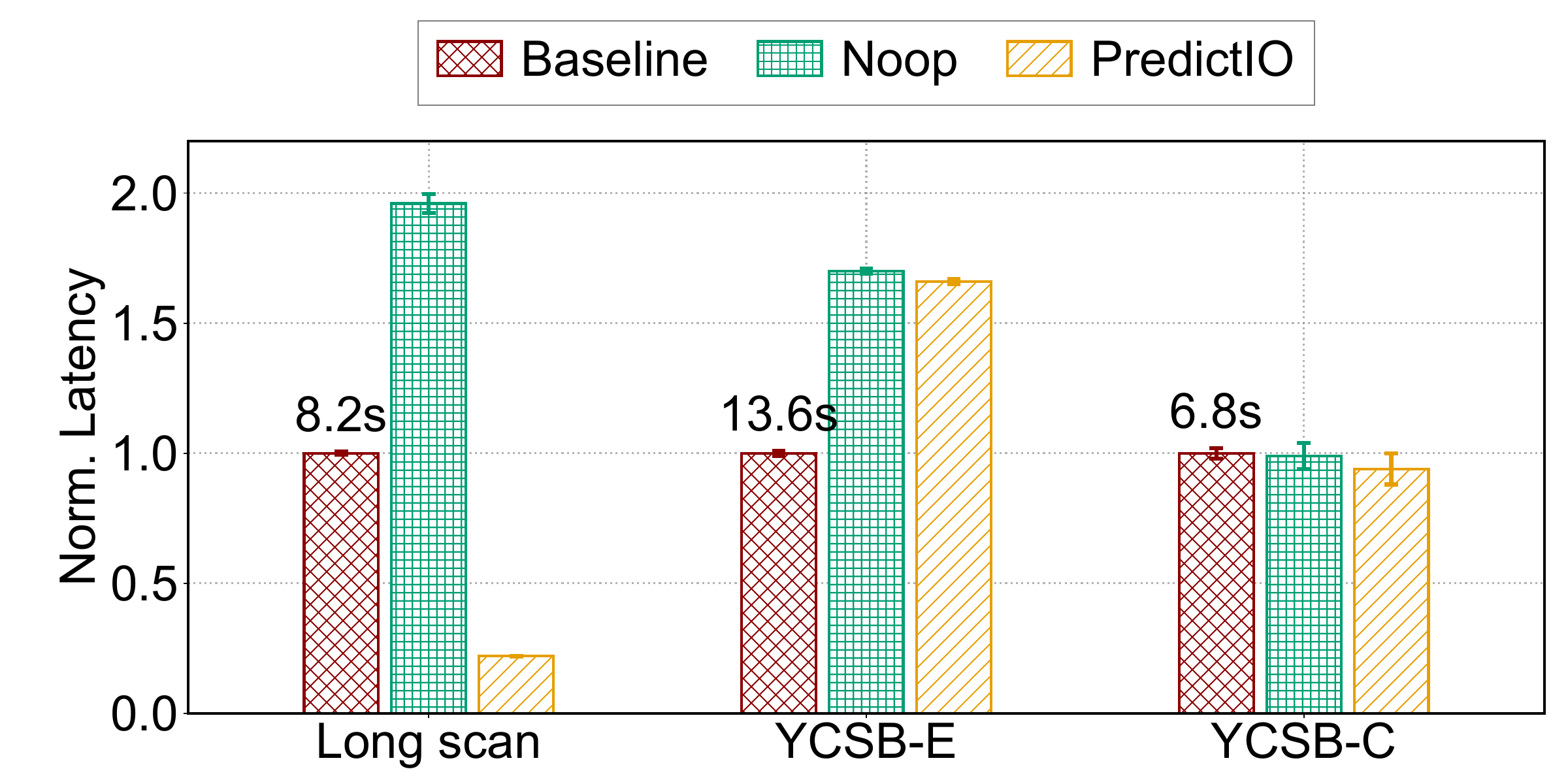}
    \vspace{-0.3cm}
    \caption{Policy migration from RocksDB into \NAME{}.}
    \label{fig:rocksdb-ycsb}
    \vspace{-0.3cm}
\end{figure}

\paragraph{Policy migration.} We evaluate whether \NAME{} can host RocksDB's native readahead policy while preserving its selectivity and benefit, as shown in Figure~\ref{fig:rocksdb-ycsb}. We test three workloads: a long sequential scan and YCSB E and C. Baseline runs RocksDB with its native readahead. The sequential scan uses a one-million-record, 1-KiB-per-record database with a 32-MiB block cache and performs a full sequential read, providing a workload type not covered by YCSB and testing whether migration preserves the policy's benefit under ideal sequential access. Workload E (95\% short scans, 5\% inserts) tests whether migration remains effective when the prefetch window is narrow. Workload C (Zipfian point reads) checks that the migrated policy preserves selectivity by avoiding unnecessary prefetching. Both YCSB workloads use one million records and 100,000 operations.

The long sequential scan benefits the most. Relative to Baseline, Noop reaches 1.96 while \NAME{} reaches 0.22, a roughly 4.6$\times$ speedup. Noop discards the migrated requests, removing RocksDB's native readahead benefit, whereas the \NAME{} system aggregates predictions into larger ranges and submits them asynchronously via the kernel backend. For YCSB-E, Noop reaches 1.70 and \NAME{} 1.66, so the migrated policy provides only a small improvement over executing without a backend. For YCSB-C, Noop and \NAME{} stay close to Baseline.

Results show policy migration is workload-dependent. The long scan is highly readahead-sensitive: Baseline uses RocksDB's native readahead, while Noop redirects hints through \NAME{} but discards requests, removing readahead benefit without replacement I/O. \NAME{} uses the same policy but aggregates predicted pages into larger ranges and submits them asynchronously via the kernel backend. With a persistent sequential stream, it accumulates enough predictions to keep pages in flight, overlapping I/O with subsequent reads and extending beyond a single native readahead window. YCSB-E's short scans start at varying positions, so the policy needs several sequential accesses to enlarge its window, and many scans finish before \NAME{} can accumulate sufficiently large requests or complete useful prefetches. It thus gains little I/O benefit while paying event and policy overhead, remaining close to Noop and well above Baseline. This gap reflects limited prediction timeliness and coverage, not steady-state runtime cost. YCSB-C has only point reads, generating no readahead events, so both Noop and \NAME{} stay close to Baseline. Overall, \NAME{} preserves the policy's selectivity, provides substantial benefit for long sequential streams, and avoids unnecessary prefetching when the policy does not apply.

\begin{figure*}[t]
\centering
\begin{minipage}[t]{0.30\textwidth}
    \centering
    \includegraphics[width=0.98\linewidth]{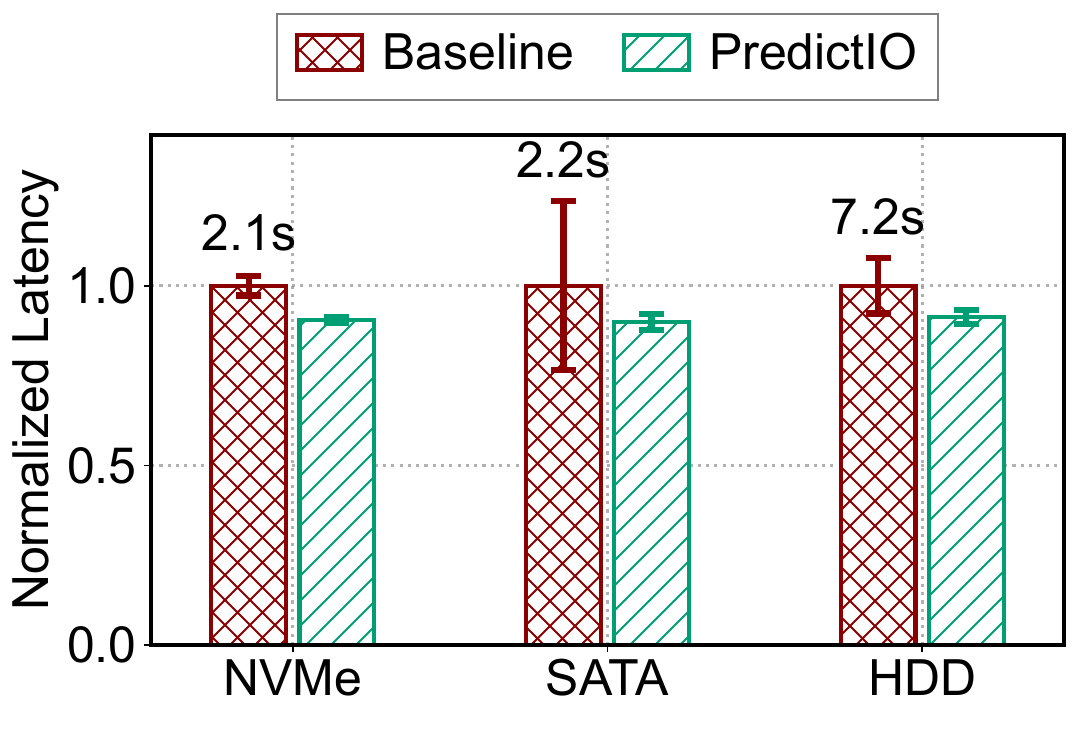}

    \vspace{-0.3cm}
    \captionof{figure}{Application launch}
    \label{fig:eval-app-launch-sens}
\end{minipage}
\hfill
\begin{minipage}[t]{0.30\textwidth}
    \centering
    \includegraphics[width=0.98\linewidth]{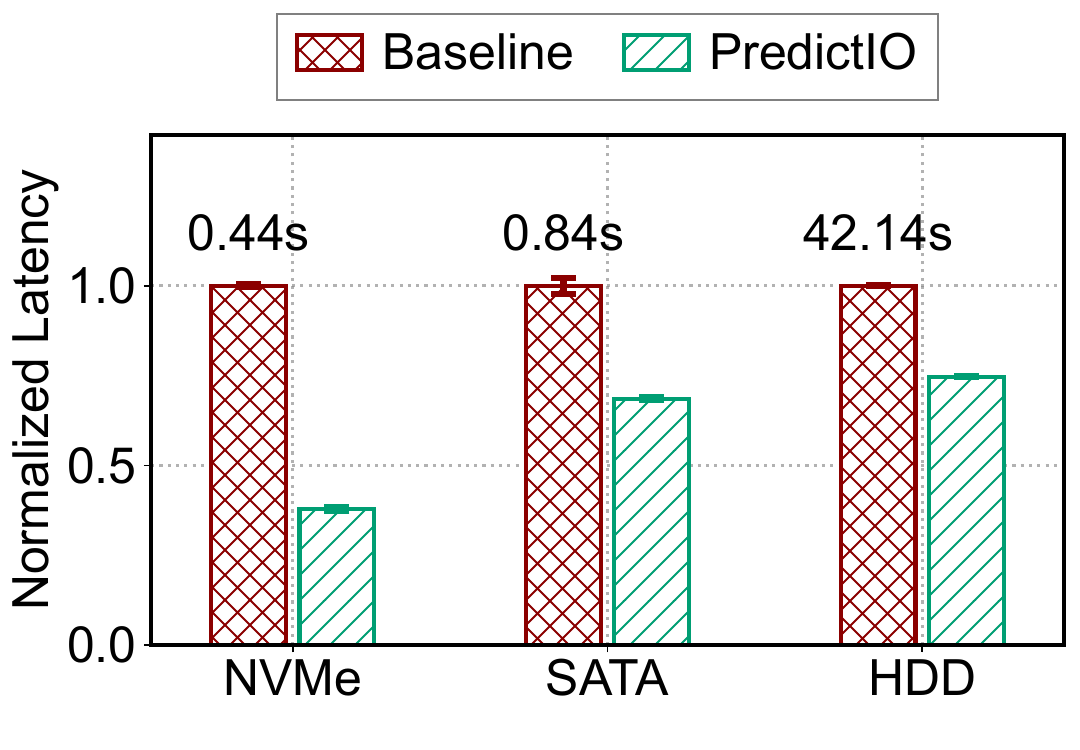}

    \vspace{-0.3cm}
    \captionof{figure}{SQLite range scan}
    \label{fig:eval-sqlite-sens}
\end{minipage}
\hfill
\begin{minipage}[t]{0.30\textwidth}
    \centering
    \includegraphics[width=0.98\linewidth]{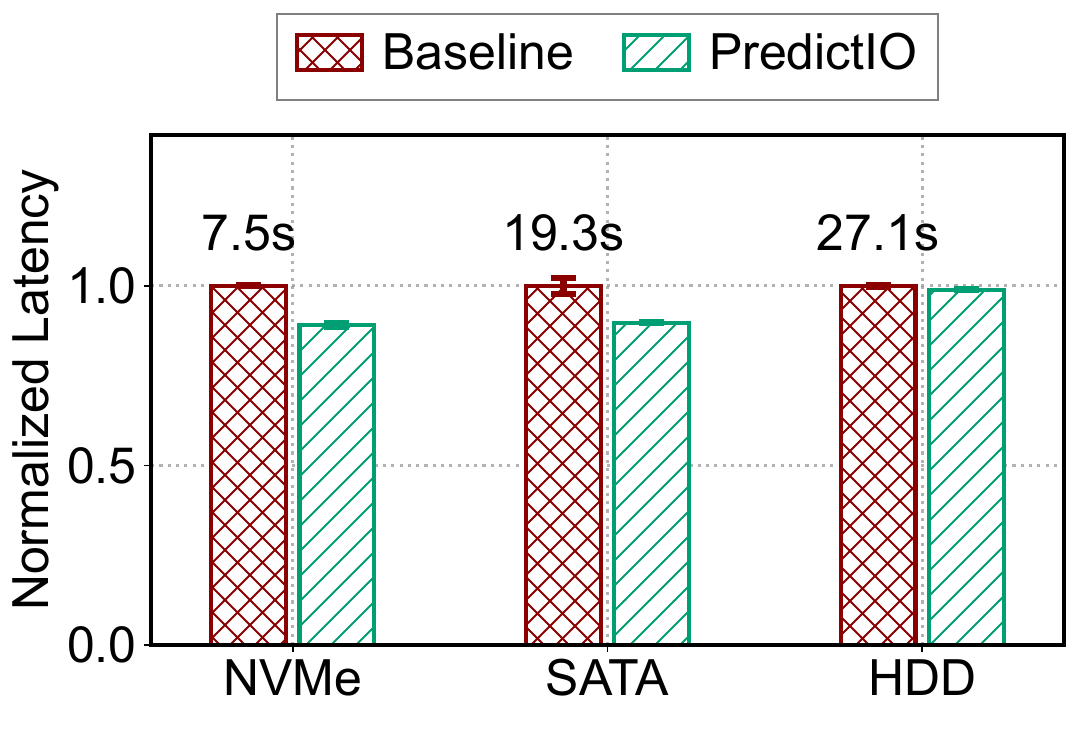}

    \vspace{-0.3cm}
    \captionof{figure}{LLM model loading}
    \label{fig:eval-llm-start-sens}
\end{minipage}
\end{figure*}

\subsection{Storage-Device Sensitivity}

To evaluate how storage characteristics affect policy delegation, we repeat representative application-launch, SQLite, and LLM experiments across an NVMe SSD, a SATA SSD, and a SATA HDD. Measured with 4\,KiB random reads using \texttt{fio}, the devices sustain approximately 1.26\,GiB/s, 240\,MiB/s, and 1.34\,MiB/s, respectively. Rather than repeating the full benchmark matrix, we select one representative configuration from each workload and compare \NAME{} against the same-device Baseline.

Application launch shows nearly identical gains, reducing Blender startup latency by 8.7--10.1\%, as shown in Figure~\ref{fig:eval-app-launch-sens}. This suggests the history-based predictor is primarily limited by prediction coverage rather than storage bandwidth. The small difference between NVMe and SATA startup times suggests much of Blender startup is not storage-bound, limiting the maximum achievable prefetching benefit. SQLite range scans show the largest improvement, achieving a 2.64$\times$ speedup on NVMe and remaining beneficial on SATA SSDs and HDDs, as shown in Figure~\ref{fig:eval-sqlite-sens}. However, the relative gain decreases as storage becomes slower. On HDDs, speculative reads compete more directly with demand reads, reducing the advantage of deeper lookahead. LLM model loading follows a similar trend: \NAME{} improves startup latency by approximately 10--11\% on both SSDs but provides little benefit on HDDs, where loading is dominated by sequential bandwidth rather than page-fault latency, as shown in Figure~\ref{fig:eval-llm-start-sens}. Across all three workloads, storage latency alone does not determine prefetch effectiveness. \NAME{} benefits most when workload semantics expose future accesses and the device can overlap speculative with demand I/O, so faster devices do not eliminate prediction's value and slower devices do not guarantee larger gains.
\section{Related Work}

\paragraph{OS-Level Prefetching.}
Prefetching hides storage latency. Early predictive prefetching~\cite{griffioen1994reducing} inferred future accesses from past references. Other work studied prefetching's interaction with cache management under known or predicted accesses: Kimbrel et al.~\cite{kimbrel1996trace} compare parallel prefetching and caching under application-provided hints, while SARC~\cite{gill2005sarc} adaptively partitions cache space between sequential and random streams. Subsequent systems target new storage and memory technologies: Leap~\cite{al2020effectively} combines majority-based online prefetching with a lightweight Linux kernel data path to improve page-cache hit rate and reduce remote latency, and SPAN~\cite{fedorov2017speculative} fetches pages for NVM storage. These systems improve prediction, scheduling, or cache management within specific OS/storage settings. Application-controlled caching~\cite{260149} advocated application involvement in cache replacement, but targeted what to keep rather than what to prefetch next.

\paragraph{Semantic-Aware Prefetching.}
Early application-directed systems expose future accesses to the OS: informed prefetching lets applications disclose them, while later designs use decoupled threads or future-access lists~\cite{patterson1995informed,yang2002decoupled,vandebogart2009reducing}. Recent systems provide richer cross-layer abstractions. CrossPrefetch~\cite{garg2024crossprefetch} splits prefetching between the OS and a user-level runtime, exposing cache state and prefetch effectiveness to guide prediction. HintStor~\cite{ge2022hintstor} provides a framework for evaluating I/O hints across heterogeneous storage. PPC~\cite{liu2026accelerating} introduces a programmable page cache for LLM loading, using I/O templates to customize prefetching and eviction. FALCON~\cite{yan2012fast} predicts mobile application launches using user context, such as location and temporal patterns, and prelaunches applications to reduce startup latency. Frog~\cite{zhang2015frog} abstracts context-specific I/O solutions as views, enabling strategies for sequential and random access patterns. Lynx~\cite{laga2016lynx} integrates Markov-chain learning into Linux readahead to adapt to SSD performance, while semantic locality~\cite{peled2015semantic} shows that application semantics capture access relationships invisible to spatio-temporal predictors.

\paragraph{User-Space Delegation in Operating Systems.}
ghOSt~\cite{humphries2021ghost} shows that delegating scheduling to user space enables rapid policy development, non-disruptive upgrades, and fault isolation without performance loss, and PPC~\cite{liu2026accelerating} applies a similar philosophy to the page cache, letting users customize cache policies for LLM loading via programmable interfaces. Skyloft~\cite{jia2024skyloft} supports microsecond-scale preemption and programmable scheduling, while PageFlex~\cite{yelam2025pageflex} uses eBPF to delegate Linux paging policies to user space with low overhead. IGTCache~\cite{wang2025efficient} unifies caching for heterogeneous AI workloads by detecting access patterns and tailoring prefetching, eviction, and allocation. Frog~\cite{zhang2015frog} abstracts context-specific I/O solutions as views, letting applications select prefetching strategies through file path strings, and ContextPrefetcher~\cite{zhang2024context} delegates prefetching management to the host for near-storage accelerators via cross-layered context abstractions. More broadly, FUSE~\cite{vangoor2017fuse} and SPDK~\cite{yang2017spdk} show that moving file system and storage logic to user space is practical for both prototyping and production use.
\section{Conclusion}

We presented \NAME{}, a programmable framework separating workload-aware prediction from kernel-controlled execution. User-space policies turn kernel-observed accesses or hints into page-range predictions, which the kernel backend filters and executes asynchronously. We instantiated \NAME{} for application launch, B-tree scans, and LLM loading, and showed that policies are reusable across storage engines or migratable from integrated designs. \NAME{} adds low control-path overhead and benefits persist across NVMe, SATA, and HDD devices. These results show that workload semantics can improve prefetching without embedding specialized policies in the kernel or moving page-cache management out of the OS.


\bibliographystyle{ACM-Reference-Format}
\bibliography{sample-base}


\end{document}